\documentstyle[preprint,aps]{revtex}
\draft
\tightenlines
\begin{document}
\setcounter{page}{0}  
%
\title{ Quantum Critical Behavior of the Infinite-range Transverse    
  Ising Spin Glass : An Exact Numerical Diagonalization Study }    
\author{Parongama Sen}
\address{Institute of Theoretical Physics,
University of Cologne,}
\address {Zulpicher Strasse 77,
50937  Cologne, Germany }
\address {e-mail: paro@thp.uni-koeln.de}

\author{Purusattam Ray}
\address{
The Institute of Mathematical Sciences, C. I. T. Campus, 
Chennai 600 113, India }
\address {e-mail: ray@imsc.ernet.in}
\author{Bikas K. Chakrabarti}
\address{Saha Institute of Nuclear Physics,
1/AF Bidhannagar, Calcutta 700064, India}
\address {e-mail: bikas@saha.ernet.in}
%
%
\maketitle
\begin{abstract}
We report {\em exact} numerical diagonalization results of the infinite-range 
Ising spin glass in a transverse field $\Gamma$ at zero temperature. 
Eigenvalues and eigenvectors are determined for various strengths of 
$\Gamma$ and for system sizes $N \le 16$. We obtain the moments 
of the distribution of the 
spin-glass order parameter, the spin-glass 
susceptibility and the mass gap at different values of $\Gamma$. 
The disorder averaging is done typically 
over 1000 configurations. Our finite size scaling analysis indicates a 
spin glass transition at $\Gamma_c \simeq 1.5 $. Our estimates for the 
exponents at the transition are in agreement with those known from 
other approaches. 
For the dynamic exponent, we get $z=2.1 \pm 0.1$ which is in contradiction with a  
recent estimate ($z=4$). Our cumulant 
analysis indicates the existence of a {\em replica symmetric} spin glass 
phase for $\Gamma < \Gamma_c$.  
\end{abstract}

\bigskip  
\noindent PACS nos : 75.10.Jm, 75.10 Nr, 64.60.Cn, 64.60.Fr 
\newpage 


Quantum phase transitions in disordered systems have been studied 
intensively in recent years \cite{bikas}. Of particular interest is the 
Ising spin glass model in a transverse field which provides a rather 
simple model where it can be shown \cite{bray-moore} that at zero 
temperature, the spin-glass ordered ground state is destabilised by 
quantum fluctuations. As a result, if one varies the strength $\Gamma$ 
of the transverse field, there can be a phase transition between the 
spin-glass ordered and disordered ground states at a critical value of 
$\Gamma = \Gamma_c$. The case of infinite range model 
is specially interesting since in the absence of the transverse field, 
the model reduces to the usual Sherrington-Kirkpatrick (SK) model 
\cite{sherrington-kirkpatrick}. The classical SK-model is very well  
studied. It has a finite temperature transition 
to a low temperature spin glass phase where replica symmetry is 
broken \cite{binder-young}. In presence of the transverse field $\Gamma$, the 
spin glass ordering occurs at lower temperatures and above a critical 
value of the field, the quantum fluctuations destroy the order. 
At non-zero temperatures, however, quantum fluctuations are unimportant in 
determining the critical behavior. In this letter, we are interested 
in the SK-model in a transverse field at zero temperature, where the   
transition between the spin glass to disordered ground state is driven   
by quantum fluctuations alone through the tuning of the strength of the 
transverse field $\Gamma$. 

There has been quite a number of studies on this model at recent times. 
The perturbation expansion of the free energy by 
Ishii and Yamamoto \cite {ishii-yama} gives the value of   
the critical field $\Gamma \simeq 1.5$ and the susceptibility exponent
$\approx 0.5$. 
The non-perturbative analysis \cite{miller-huse}, the mean-field theory 
of quantum rotors \cite{sachdev} and the numerical techniques such as 
quantum Monte Carlo method \cite{alvarez-ritort} give consistent 
estimates of the critical field and the critical exponents except 
the value of the dynamic exponent $z$. The 
Monte Carlo study suggests the value of $z = 4$, much 
too higher than the value $z=2$ obtained in other studies. 
In a recent work, the Schr\" odinger equation for the model has been solved 
numerically \cite{lancaster-ritort} and the interaction energy and 
longitudinal susceptibility are determined. The estimates of the critical 
field and the exponents from this study match with the analytical 
results. But, in this method the exponents are obtained not directly 
individually, but rather through a scaling relation. 
An important observation that comes out of the mean field theory 
is the possibility of a replica symmetric spin glass phase in quantum 
transition below $\Gamma << \Gamma_c$ \cite{ray,sachdev}. This point 
certainly deserves further attention. It may be noted that most of these
results are obtained using the approximate classical mapping of the 
quantum systems.

Here we report the results of exact diagonalization of the infinite range 
Ising spin glass model in transverse field,  using the Lanczos 
algorithm for system sizes $N \le 16$. Configuration 
averaging is done typically over 1000 realizations of bond values (for 
low $N$). We obtain the various moments of the 
order parameter distribution $P(q)$  
and the mass gap. The cumulant analysis of $P(q)$ provides an estimate 
of the critical field $\Gamma_c \simeq 1.5$ in the limit of large $N$. 
This value of $\Gamma_c$ is comparable to the values obtained from other 
studies. For spin glass susceptibility and the mass gap, our finite 
size scaling analysis follows the idea of the phenomenological 
renormalization group transformation \cite{barber}. Our estimate of 
the correlation length exponent $\nu = 0.252 \pm 0.004 $, the susceptibility exponent 
$\gamma = 0.51 \pm 0.02$ and the dynamic exponent $z= 2.1 \pm 0.1 $
(obtained from the scaling 
of the mass gap \cite{hammer}) compares well with the values obtained 
from other studies.  Our estimate of $z$ does not match with the quantum 
Monte Carlo result \cite{alvarez-ritort} which we think is too high and 
may be an artifact of the size effect along the Trotter direction. Our 
cumulant analysis shows that for $\Gamma << \Gamma_c$,  
$P(q)$ attains a two peak structure (related by the 
spin inversion symmetry), the peaks becoming narrower with $N$. This is 
similar to what one gets in the classical SK-model above the 
Almeida-Thouless line (see \cite{binder-young}) where the glass phase 
is replica symmetric. The extended width  of $P(q)$,
characteristic of replica broken phase, is clearly absent in our study.  



The SK model in a transverse field is described by the  Hamiltonian 
$$H = \sum_{<ij>} ^N J_{ij}S_{i}^{z}S_{j}^{z} - \Gamma \sum_{i}^N S_{i}^{x}
\eqno (1) $$
where the $J_{ij}$'s are long ranged and follow the Gaussian distribution
$$D(J_{ij}) = ({N\over{2\pi J^2}})^{1/2} \exp({-NJ_{ij}^2\over {2J^2}}) 
$$  
and $<ij>$ in the summation denotes that each pair of spins is taken
only once.  $S_{i}^z$ and $S_{i}^x$ are Pauli spin matrices. 

We have performed exact diagonalisation of the Hamiltonian (1) 
 using  Lanczos algorithm for  $N$ = 4, 6, 8, 10, 12, 14 and 16
and repeated for various realisations of   $J_{ij}$ values. For each   
realization of $J_{ij}$, all the relevant quantities like the 
order parameter etc. are computed and then averaged over all the  
realizations. For the diagonalisation precedure, the 
 basis states $|\phi_\alpha >$ are chosen to be the eigenstates
of the spin operators $\{S_{i}^z, i= 1,N\}$. 
In the absence of the transverse field term, the Hamiltonian (1) reduces
to the Hamiltonian of the S-K model and is diagonal in this representation.
With the transverse field present, each  eigenstate of (1) is   obtained as
 a  superposition of the basis states 
which are $2^N$ in number  (corresponding to $S_{i}^z = \pm 1$).
The  $n$th eigenstate for the quantum Hamiltonian is  written as 
$$|\psi_n> = \sum_{\alpha=1}^{2^N} a_{\alpha}^n |\phi_\alpha>$$
and the ground state is denoted by $|\psi_0>$.

We  calculate the moments of the distribution $P(q)$   
of the order parameter $q = (1/N) \sum_i \overline{ <S_i^z>^2}$.
The overhead bar indicates the configuration average and the $< ... >$ denotes
the expectation values at the ground state. 
$P(q)$ is
not directly obtained here, rather
the moments $m_k= \overline {q_k} = \int_0^1 q^k P(q) dq$ 
are calculated using \cite{binder-young},

$$ q_k = {1\over {N^k}} \sum^{N}_{i_1}....
\sum^{N}_{i_k}<S_{i_1}^z...S_{i_k}^z>^2. 
\eqno (2)$$
It can be easily shown that the RHS of the above equation  gives the 
$k$th moment of the distribution. 
We take for example

$$q_2 = 
{1\over {N^2}}\sum^{N} _{i_1}\sum^{N} _{i_2}[<S^{z}_{i_1}S^{z}_{i_2}>
<S^{z}_{i_1}S^{z}_{i_2}>]$$
Now, $<S^{z}_{i_1}S^{z}_{i_2}>$ in the ground state is given by 
$<\psi_0|S^{z}_{i_1}S^{z}_{i_2}|\psi_0>$. 
The latter quantity is again  given in terms 
of the basis  states,  i.e., 
$<\psi_0|S^{z}_{i_1}S^{z}_{i_2}|\psi_0> = \sum_{\alpha}^{2^N}
|a_{\alpha}^0|^2 < S_{i_1}^z S_{i_2}^z>_{\alpha}$, where $< ... >_{\alpha}$
denotes the product of the $i_1$th and $i_2$th spin in the $\alpha$th
basis state.
Writing $|a^{0}_\alpha |^2$ = $\omega _\alpha $, the RHS of (2) is
therefore given by

$$ = {1\over {N^2}}\sum^{N} _{i_1}\sum^{N} _{i_2}
[\sum_{\alpha}^{2^N} 
\omega_\alpha 
<S^{z}_{i_1}S^{z}_{i_2}>_{\alpha }
\sum ^{2^N}_{\beta} \omega_\beta  <S^{z}_{i_1}S^{z}_{i_2}>_{\beta }].$$


\noindent After configurational averaging, it gives
$m_2 = \int dq P_J(q)q^2.$

The exact diagonalisation procedure gives us the 
$\omega _\alpha $'s. Using the above,  we have calculated the 
Binder cumulant \cite{binder} for a value of $N$ and $\Gamma$ as    
$$g_N(\Gamma) = \frac{1}{2}[3 - \frac{m_4}{(m_2)^2} ] \eqno(3)$$  
and the   nonlinear susceptibility  
$$\chi = (1/N) \sum_{ij}^{N} <S^{z}_iS^{z}_j>^2. \eqno(4) $$
We calculate the order parameter and the spin glass susceptibility which 
are given by the first and second moments of the distribution $P(q)$.  
While calculating the moments of $q$, we take a set of  half of the
basis states, ignoring the     
other half, which consists of the trivially degenerate ones and obtained 
by simply flipping
the spins in the states of this  set.
This is necessary as otherwise, even in  the ferromagnetic case
one will get $<S_{i}^z> = 0$, as the two degenerate states are equally
probable. The quantities have to be suitably   rescaled to get the proper 
values.  

The  energy eigenvalues of  the $n$th state is  given by 
$E_n =  <\psi_n| H |\psi_n>$.  
The mass gap is  
$$\Delta E  = E_1 - E_0 \eqno(4)$$ 
where $E_0$ is the 
ground state energy and $E_1$ is the energy of the 
first excited state. We determine the average mass gap $\Delta E_N(\Gamma)$ 
for different values of of $\Gamma$ and $N$.  





We estimate first the cooperative energy per spin, which is the
expectation value of the cooperative part of the 
Hamiltonian in the ground state (in this case, it is simply obtained by 
operating the  first term of the Hamiltonian (1) on $|\psi_0>$). 
The asymptotic value of this quantity, which gives the classical 
ground state energy per spin in the S-K model (in the limit  
$\Gamma \rightarrow 0 $), is found to 
be equal to $-0.74 \pm 0.01$ (in units of $J$). This  
agrees very well with the ground state energy ($-0.76)$ of the
classical SK model \cite{binder-young}.

The Binder cumulant $g_N(\Gamma)$ as a function of $\Gamma$ is shown   
in Fig. 1 for different values of $N$.    
The form of $g_N(\Gamma)$ changes with the system size 
and near $\Gamma_c$ is expected to scale with $N$ in the 
following way \cite{alvarez-ritort}
$$g_N(\Gamma) \sim f((\Gamma - \Gamma_c) N^x) \eqno(5)$$ where 
$f(X)$ is a scaling function and $x$ is related to the mean field 
correlation length exponent $\nu$. $g_N(\Gamma)$ versus $\Gamma$ for various $N$ 
then yield a family of curves. These curves  intersect at a common point 
which gives the value of $\Gamma_c$. 
Due to the corrections to finite size scaling, the intersections of 
the $g_N(\Gamma)$ curves for all $N$ values may not coincide at a common point 
(see \cite{binder-young} for discussion). We determine the point of 
intersection $\Gamma_c(N,N')$ of the $g(\Gamma)$ curves for two system sizes 
$N$ and $N'$ and plot it against $1/(NN')^{1/2}$ in Fig. 2. The value of 
$\Gamma_c(N,N')$ converges to $\Gamma_c \sim 1.5 \pm 0.1$ in the limit of 
$N, N' \rightarrow \infty$. This value of $\Gamma_c$ is comparable to the 
value obtained from the non-perturbative analysis \cite{miller-huse}, 
the Pad\'e treatment \cite{yamamoto-ishii} and the solution of the 
Schr\"odinger equation \cite{lancaster-ritort}. 

The finite size scaling concept of Fisher and Barber extended to 
the long-ranged systems suggests that the exponent $x = 1/(\nu_{mf} d_c)$, 
where $\nu_{mf}$ is the mean field correlation length exponent and $d_c$ 
is the upper critical dimensionality of the corresponding short range 
system \cite{botet}. For any pair of systems with total spins $N$ and 
$N'$, we first find out the point of intersection $\Gamma_c(N,N')$. 
With $\Gamma_c = \Gamma_c(N,N')$ we try to scale $g_N(q)$ and $g_{N'}(q)$ 
around $\Gamma_c$ with a suitable value of $x$, so that all the 
data points fall on the same scaling curve $f(X)$. Fig. 3 
shows the value of $x$ against $1/\sqrt {NN'}$.   
The value of $x$ is almost independent 
of $N$ and we get $x = 0.66 \pm 0.01$. If we take $d_c = 6$ which is 
the upper critical dimension of the classical Ising spin glass we 
get $\nu = \nu_{mf} = 0.252 \pm 0.004$ in agreement with the 
predicted value of $\nu = 1/4$ \cite{sachdev,alvarez-ritort}. 

The finite size scaling form for the spin glass susceptibility $\chi$ 
can be written as 
$$\chi_N(\Gamma) \sim N^y \Phi((\Gamma - \Gamma_c)N^x) \eqno(6)$$ 
where the exponent $y = x\gamma$, $\gamma$ being the exponent which 
characterizes the divergence of the susceptibility at $\Gamma_c$ 
in the thermodynamic limit. $\Phi(X)$ is a 
scaling function. The finite size analysis can be done invoking the idea of 
the phenomenological renormalization group transformation (see  
\cite{barber} for a review). If $\Phi(X)$ is a power function 
of $X$, then the above scaling form is 
satisfied exactly for any two finite systems of spin numbers $N$ and $N'$ 
in the sense that $\chi_N(\Gamma)/N^y = 
\chi_{N'}(\Gamma')/N'^y$ with the recursion 
relation $\Gamma - \Gamma_c = 
(\Gamma' - \Gamma_c)(N'/N)^x$. The fixed point $\Gamma^*$ of the 
transformation satisfies $$ \frac{\chi_N(\Gamma^*)}{\chi_{N'}(\Gamma^*)}
= \left (\frac{N}{N'}\right)^y,$$ 
where $\Gamma^* \rightarrow \Gamma_c$ as $N, N'  
\rightarrow \infty$. For any two sizes $N$ and $N'$, we take   
$\Gamma^*$ from the intersection of the curves 
$g_N(\Gamma)$ and $g_{N'}(\Gamma)$ 
and apply the above fixed point equation to get the value of $y$. The 
resulting estimate of $y$ depends on $N$ and $N'$ and approaches the 
exact value as $N, N' \rightarrow \infty$ (see \cite{dossantos} for 
discussion). Fig. 4 shows the value of $y$ plotted against $1/N_m$, where
$N_m = \sqrt(NN')$, from 
which we determine the asymptotic value of $y = 0.33 \pm 0.01$. From $y$ 
and $x$, we get $\gamma = 0.51$ which agrees with the value 1/2 obtained 
in other studies \cite{sachdev,alvarez-ritort}.  

The scaling form for the mass gap $\Delta E_N(\Gamma)$    
can be written as \cite{hammer}
$$\Delta E_N(\Gamma) \sim N^{1-z} \Psi ((\Gamma-\Gamma_c) N^x)$$ where 
$\Psi$ is a scaling function. Employing the same scaling analysis as above, 
for the  mass gap, we obtain the value of the exponent $z$. 
Fig. 5 shows the value of $z$ plotted 
against $1/N_m$. We find the asymptotic value of value $z = 2.1 \pm 0.1$.
This value of $z$ agrees with its predicted value in \cite{sachdev} but 
disagrees with the quantum Monte Carlo simulation result $z=4$.   

Proceeding in the same way, we find $\beta = 1.0 \pm 0.1$ where 
$\beta $ is the order parameter exponent, i.e., in the 
thermodynamic limit  $q \sim (\Gamma - \Gamma_c)^\beta .$
This is  in agreement with  the previous studies \cite{bikas}.

>From Fig. 1, we find that the behavior of the Binder cumulant is very 
similar to what happens in normal spin system. If $g(T)$ is the value 
of the cumulant in the thermodynamic limit at any temperature $T$, then 
for a normal ferromagnetic system it is expected that $g(T) = 1$ for 
$T << T_c$ and $g(T)=0$ for $T >> T_c$. The value of $g(T_C)$ at 
$T_C$ depends on the dimensionality of the system. At low $T$-values, 
the value attained by the Binder cumulant depends on $N$ and extrapolates 
to the value unity in the thermodynamic limit. This coincides with the 
fact that the order parameter distribution ($P(m)$ of magnetisation $m$ for 
a ferromagnet) reduces to a delta function (modulo symmetric operation) 
in the limit $N \rightarrow \infty$ and the system goes to a definite 
symmetry broken state. 

This is not the case where replica symmetry is broken like in classical 
SK model at low temperatures and zero external field or spin glass models 
in higher dimensions. Here $P(q)$ does not reduce to a delta function in 
the ordered phase even at $N \rightarrow \infty$. Instead, it consists 
of a delta function plus a continuous, almost size independent part 
going right down to $q=0$. As a result, in the spin-glass phase, the 
Binder cumulant attains a value which is significantly lower than unity 
and does not show any change with the system size \cite{marinari}.   

Our $g(\Gamma)$ curves in Fig. 1 for different 
$N$ values show a crossing point (corresponding to a phase transition) 
at a finite value of $\Gamma = \Gamma_c$, below which the curves splay 
out for different $N$ and then saturate to values which tend to unity 
with increasing $N$. This suggests that for $\Gamma << \Gamma_c$, the 
system goes to a definite replica symmetric spin-glass ordered ground state.

\medskip

In summary,  all our estimates for the critical tunnelling field (critical
point) and exponents agree with the previous estimates, our estimate for
$z$ agrees with that estimated by Miller and Huse 
\cite{miller-huse} and also Ye et al \cite {sachdev}, while it disagrees
that obtained by Alvarez and Ritort \cite{alvarez-ritort}. 
We believe, the high value of $z$ obtained
in the quantum Monte Carlo study (of Ritort
et al) is due to an artifact of the
size anisotropy in the Trotter direction. We also emphasize that the
variation of $g_N(\Gamma)$ functions with $N$ for small
$\Gamma$ indicates the quantum
spin glass phase (for $\Gamma$ below $\Gamma_c$) is replica 
symmetric in this model.
\medskip

\noindent We are grateful to M. Acharyya,   
 S. M. Bhattacharjee, A. Dutta,  H. Rieger and  R. R. dos Santos
 for many discussions and suggestions. 
We also thank D. Dhar for a careful reading
of the manuscriot and comments. PS acknowledges 
support from SFB 341.

\pagebreak


\vskip 1.0 true cm

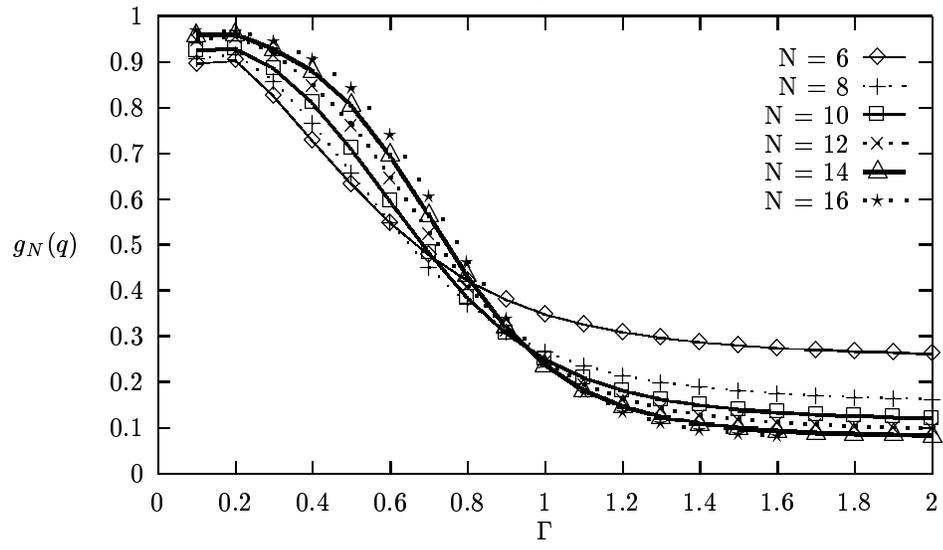
\begin{figure}

\caption  { The variation of $g_N(\Gamma)$ is shown against the transverse
 field $\Gamma$ for different system sizes $N$. Note that the 
intersection of the curves shift towards larger $\Gamma $ values as $N$
is increased.
 }
\end {figure}

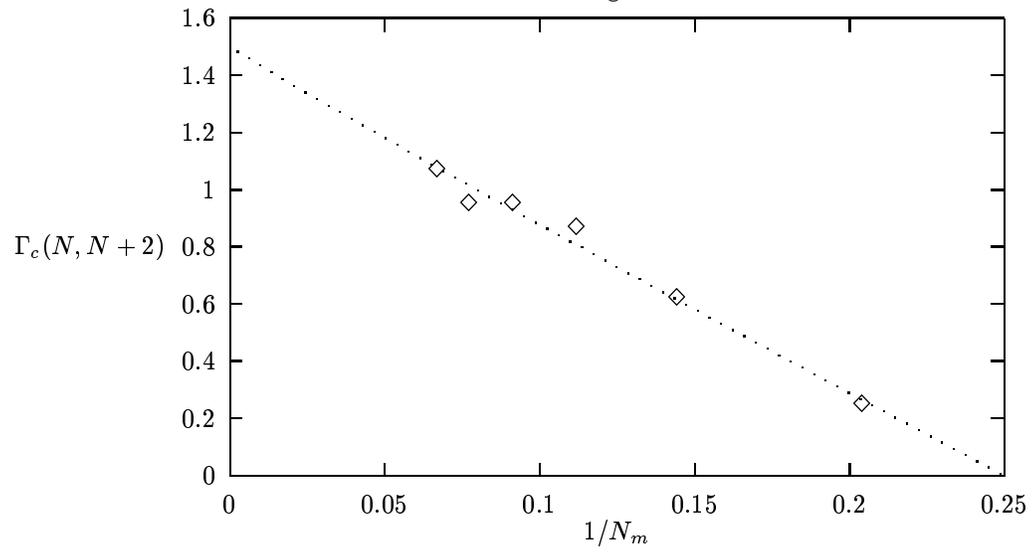
\begin{figure}

\caption 
 {The effective critical transverse field $\Gamma_c(N, N+2)$ against  
$1/N_m = 1/\sqrt{N(N+2)}$ are shown. $\Gamma_c $ for the infinite  
system is found to be at 1.5.
The best fit line is shown.   
 }
\end {figure}
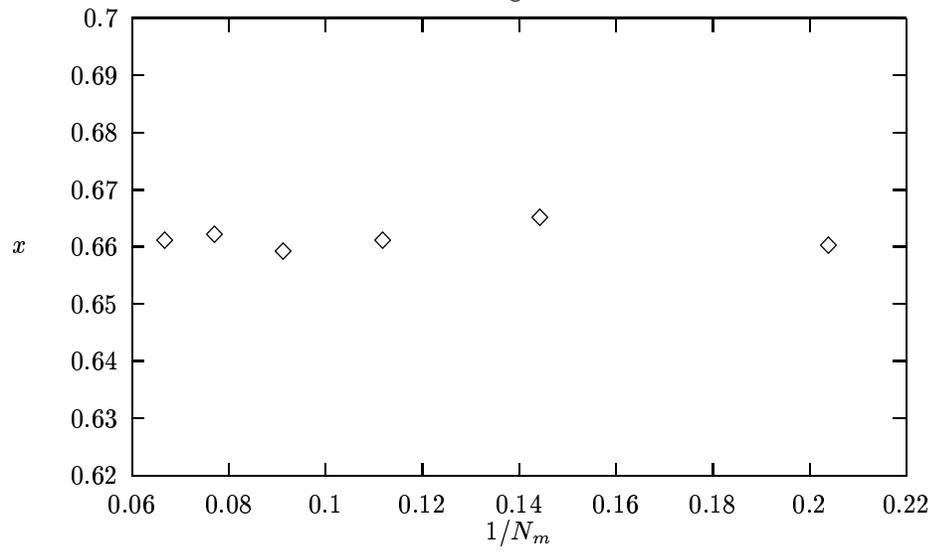
\begin{figure}
\caption  {The effective values of $x$  against  
$1/N_m = 1/\sqrt{N(N+2)} $ are shown to be fairly independent  
of the system sizes. 
}
\end{figure}

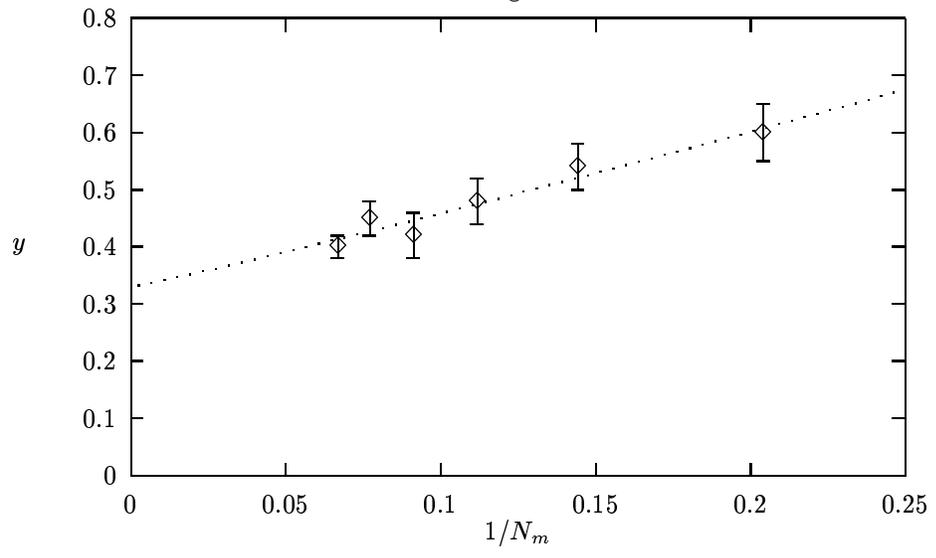
\begin{figure}
\caption  {The effective values of  $y$ 
are shown against $1/N_m = 1/\sqrt{N(N+2)}.$ 
 The best fit curve with an asymptotic 
value $y$ = 0.33 is also shown.
 }
\end{figure}
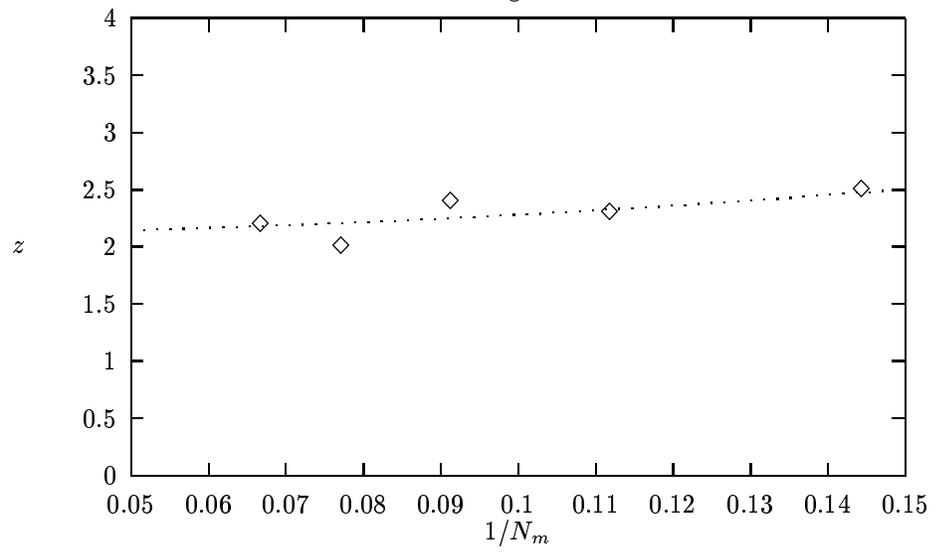
\begin{figure}
\caption   {The effective values of $z$  against 
$1/N_m = 1/\sqrt{N(N+2)}$ 
are shown.
 The best fit curve with an asymptotic 
value $z$ = 2.1 is also shown.
}
\end{figure}



\end{document}